\documentclass[aps,prstab,preprint,groupedaddress,showpacs,amsmath,amssymb]{revtex4-1}
\usepackage{epsfig}
\usepackage{dcolumn}
\usepackage{bm}

\newcommand{\Htp}{H$_{\mathrm{3}}^{\mathrm{+}}$}
\newcommand{\degs}{$^{\circ}$}
\newcommand{\Ul}{$^{\mathrm{238}}$U$^{\mathrm{27+}}$}
\newcommand{\Uh}{$^{\mathrm{238}}$U$^{\mathrm{73+}}$}
\newcommand{\Oo}{$^{\mathrm{16}}$O$^{\mathrm{3+}}$}

\begin{document}


\title{Concept for controlled transverse emittance transfer within a linac ion beam}


\author{L.~Groening}
\affiliation{GSI Helmholtzzentrum f\"ur Schwerionenforschung GmbH, Planckstrasse
~1, D-64291~Darmstadt, Germany}


\date{\today}

\begin{abstract}
For injection of beams into circular machines with different horizontal and
vertical emittance acceptance, the injection efficiency can be increased if
these beams are flat, i.e. if they feature unequal transverse emittances.
Generation of flat electron beams is well-known and has been demonstrated
already in beam experiments. It was proposed also for ion beams that were generated
in an Electron-Cyclotron-Resonance (ECR) source. We introduce an extension of the
method to beams that underwent charge state stripping without requiring their
generation inside an ECR source. Results from multi-particle simulations are
presented to demonstrate the validity of the method.
\end{abstract}

\pacs{41.75.Ak, 41.85.Ct, 41.85.Ja}

\maketitle


Beams provided by linacs generally have equal or quite similar horizontal and vertical
emittances ($\epsilon_x$ and~$\epsilon_y$). If beams from linacs are to be
injected into a subsequent circular machine, the injection process might impose
different acceptance limits ($A_x<A_y$) on the two transverse emittances. One
common injection scheme is the multi-turn injection (MTI) using a time dependent orbit
bump in one plane for beam stacking in this plane. As a consequence the injected beam
emittance in this plane, the horizontal for instance, must be significantly lower with
respect to the other one. To meet the acceptance criteria in both planes, linac
designers and operators try to keep both transverse emittances below the more stringent
acceptance limit:
\begin{equation}
\epsilon_x \cdot \epsilon_y \,=\,\epsilon_x^2\,<\,A_x^2\,.
\end{equation}
The requirement to the product of the emittances is relaxed if the linac
beam will feature different transverse emittances as well, i.e. emittances being
adopted to the acceptances of the machine into which injection has to be performed:
\begin{equation}
A_x^2\,<\,\epsilon_x \cdot \epsilon_y \,<\,A_x \cdot A_y\,.
\end{equation}
To this end a single pass beam line is needed that provides transformation of a round
uncorrelated beam
($\epsilon_x=\epsilon_y$) to a flat beam ($\epsilon_x<\epsilon_y$). Although for
electrons round to flat transfer was proposed~\cite{PhysRevSTAB.4.053501} and has been
demonstrated experimentally~\cite{MOB13_LINAC2000}, to our knowledge it has not been
proposed for initially uncorrelated round ion beams. Such an ion beam line must be
different from an emittance swapping beam line that exchanges the two different
transverse emittances by using three skew quadrupoles for
instance~\cite{PhysRevE.66.016503}. 

\section{Required Mathematical Tools}
A round to flat transformation implies beam line elements that cause coupling of
the horizontal and vertical plane. It needs a four dimensional description.
This section introduces the quantities referred to in the paper.
The next section deals with a conceptual layout of a
round to flat transformation for ions extracted from an ECR
(Electron-Cyclotron-Resonance) source. Finally, a
concept is presented that provides round to flat transformation for any uncorrelated
ion beam being charge state stripped during transportation. GSI aims at experimental
verification of the concepts. Multi-particle simulations were done
for sections planned to be integrated into the existing Universal Linear
Accelerator (UNILAC) at GSI.
\\
Any round to flat operation should only use beam line elements being
linear in the four dimensional transverse phase space. In that case the transportation
$M$ of single particle coordinates from a position $1$ to a position~$2$ is
\begin{equation}
\label{transport_matrix}
\begin{bmatrix}x \\ x' \\ y \\ y'\end{bmatrix}_2\,=\,
\begin{bmatrix}m_{11} & m_{12} & m_{13} & m_{14}\\
m_{21} & m_{22} & m_{23} & m_{24}\\
m_{31} & m_{32} & m_{33} & m_{34}\\
m_{41} & m_{42} & m_{43} & m_{44}\\
\end{bmatrix}
\begin{bmatrix}x \\ x' \\ y \\ y'\end{bmatrix}_1.
\end{equation}
As coordinates we use ($x,y$) to indicate transverse displacements and ($x',y'$) to
indicate their derivatives w.r.t. the position $s$ along the design orbit.
In the following sections we refer to beam dynamics simulation results from a code
that also uses these coordinates. Additionally, the experimental proof-of-principles
will be based on measurements of these coordinates. For this reasons we do not use the
conjugate coordinates that additionally include the contribution of the magnetic
vector field $\vec{A}$ in the momenta.
\\
The linear transport preserves the four dimensional rms emittance defined through the
beam's second moment matrix
 \begin{equation}
\label{M_Moments}
C\,=\,\begin{bmatrix}<xx> & <xx'> & <xy> & <xy'>\\
<x'x> & <x'x'> & <x'y> & <x'y'>\\
<yx> & <yx'> & <yy> & <yy'>\\
<y'x> & <y'x'> & <y'y> & <y'y'>\\
\end{bmatrix}
\end{equation}
and
\begin{equation}
\label{eps_4d}
\epsilon_{4d}^2\,=\,det\,C.
\end{equation}
Second moments of the beam are transported using the matrix equation
\begin{equation}
\label{def_transp_moments}
C_2\,=\,M\,C_1\,M^T.
\end{equation}
Coupling between horizontal and vertical plane results in
\begin{equation}
\label{eps4d_biggerorequal}
\epsilon_x\cdot \epsilon_y\,\ge\,\epsilon_{4d}
\end{equation}
with equality just for zero inter-plane coupling moments. Here
$\epsilon_x$ and~$\epsilon_y$ are the rms emittances defined through the beam moments
in their sub phase spaces in an analogue way to Equ.~\ref{eps_4d}.
Assuming an initial round beam without any inter-plane correlations, the transformation
into a flat beam cannot be accomplished by applying symplectic transformations
only~\cite{Brown_SlacRep1989}. Symplectic transformations $M$ preserve $\epsilon_{4d}$,
meet the condition
\begin{equation}
\label{def_symplectic}
M^TJM\,=\,J
\end{equation}
with
\begin{equation}
\label{J_matrix}
J\,:=\,
\begin{bmatrix}0 & 1 & 0 & 0 \\
-1 & 0 & 0 & 0\\
0 & 0 & 0 & 1\\
0 & 0 & -1 & 0\\
\end{bmatrix},
\end{equation}
and may just increase both transverse emittances ($\epsilon_x$,~$\epsilon_y$)
while keeping them equal to each other.
\\
Rigorous analysis of coupled four dimensional transverse beam dynamics can be found
in~\cite{Edwards_Teng,Willeke,Lebedev}. This section is concluded in providing eidetic
justification
for the beam line elements, which a round to flat transformation should include.
In order to achieve emittance reduction in one plane, the horizontal for instance,
a non-symplectic transformation $M_{ns}$ must be integrated into the round to flat
transformation section. This transformation anyway should be linear to preserve
$\epsilon_{4d}$. The section must also include an element $M_t$ that
decreases the horizontal emittance and increases the vertical one. This element might
act as
\begin{eqnarray}
\label{friction}
\delta x'\, & = & -a\cdot x' \\
\delta y'\, & = & \,\,\,\,\,b\cdot y',\,\,\,\,\,\,\,\,(a,b)\,>\,0\,, \nonumber
\end{eqnarray}
where a thin element not changing the particle position is assumed. The element
imposes an effective linear friction in the horizontal plane and a linear heating in
the vertical plane. If at the entrance to element $M_t$ the distribution has large
inter-plane coupling moments \mbox{$<\!x'y\!>$} and \mbox{$<\!xy'\!>$} imposed by
preceding elements, the approximation
\begin{eqnarray}
x'\, & \approx & \frac{<x'^2>}{<x'y>}\,y \\
y'\, & \approx & \frac{<y'^2>}{<xy'>}\,x \nonumber
\end{eqnarray}
turns Eq.~\ref{friction} to
\begin{eqnarray}
\label{pseudo_skew}
\delta x'\, & \approx & -a\,\frac{<x'^2>}{<x'y>}\,y\,:=\,q_s\cdot y \\
\delta y'\, & \approx & \,\,\,\,\,b\,\frac{<y'^2>}{<xy'>}\,x\,:=\,q_s\cdot x \nonumber
\end{eqnarray}
being the transformation of thin skew quadrupole with focusing strength
\begin{equation}
q_s\,=\,-a\frac{<x'^2>}{<x'y>}
\end{equation}
and
\begin{equation}
\frac{a}{b}\,=\,-\,\frac{<x'y>}{<xy'>}\cdot\frac{<y'^2>}{<x'^2>}\,,
\end{equation}
implying that \mbox{$<\!x'y\!>$} and \mbox{$<\!xy'\!>$} need to differ in sign.
\\
The required condition of different signs of the moments \mbox{$<\!x'y\!>$} and
\mbox{$<\!xy'\!>$} at the skew quadrupole entrance can be achieved by including
a transformation
\begin{equation}
\label{SolFringe}
M_{SolFringe}\,=\,
\begin{bmatrix}1 & 0 & 0 & 0 \\
0 & 1 & +K & 0\\
0 & 0 & 1 & 0\\
-K & 0 & 0 & 1\\
\end{bmatrix}
\end{equation}
through a solenoid fringe field into the beam line preceding the skew quadrupole with
\begin{equation}
\label{def_KSol}
K\,:=\,\frac{B}{2(B\rho)}\,,
\end{equation}
$B$ as the solenoid on-axis magnetic field strength, and $(B\rho)$ as the beam
rigidity. The $K$ values at the entrance and exit fringe are equal in
magnitude but they differ in sign. Solenoid fringe fields can provide the
non-symplectic transformation $M_{ns}$ that preserves $\epsilon_{4d}$. Summarizing,
a round to flat transformation of an initial beam without inter-plane correlations can
be achieved if it includes solenoidal fringe fields causing a non-symplectic
transformation
$M_{ns}$ and skew quadrupoles for the emittance transfer. This does not mean that the
transformation can be achieved using just two elements: a fringe field and a single
thin skew quadrupole. Both elements are essential ingredients but they are not
sufficient. Apart from this it is physically not possible to create a stand alone
fringe field since static magnetic field lines are closed, i.e. an entrance
fringe field intrinsically causes an exit fringe field.

\section{Beams extracted from an ECR Source}
\label{ecr_beam}
\begin{figure}
\centering
\epsfig{file=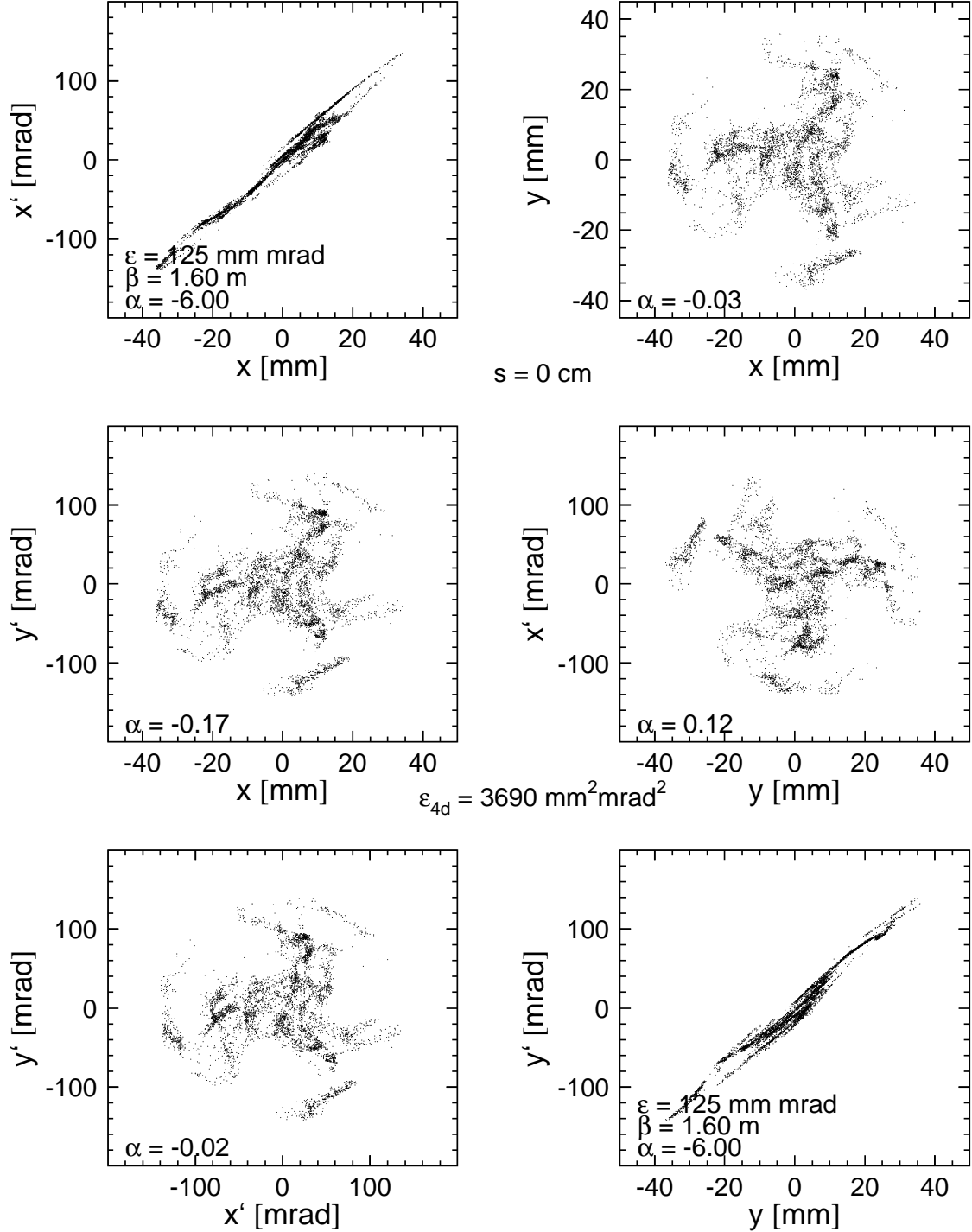,width=15cm,clip=}
\caption{Two dimensional projections of the ECR transverse phase space
distribution at the entrance to the emittance transfer section. Horizontal
and vertical rms Twiss parameters are indicated as well as the four
inter-plane correlation parameters.}
\label{fig_tphell_ecr_s1}
\end{figure}

\begin{figure}
\centering
\epsfig{file=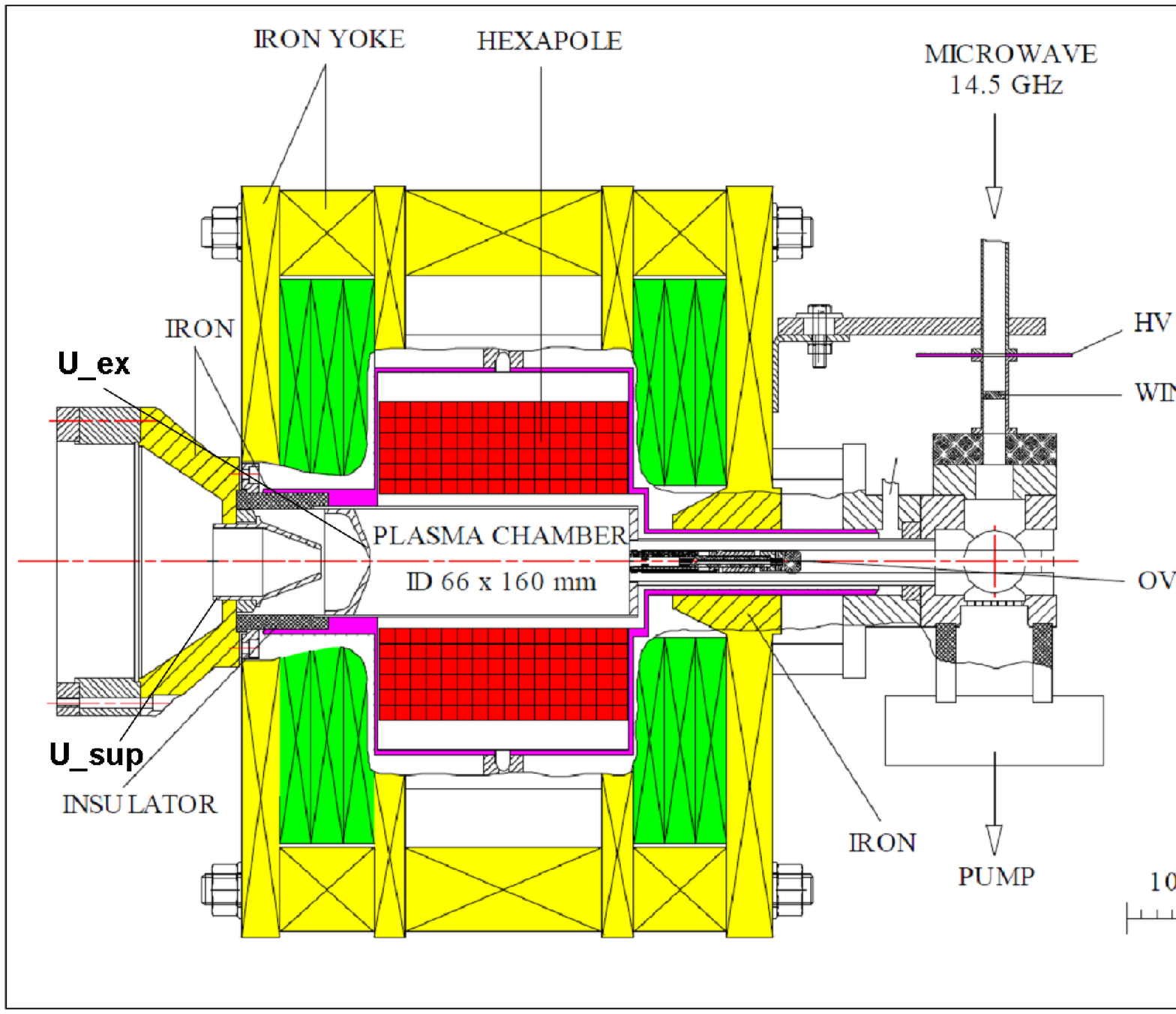,width=11cm,clip=}
\caption{(Color) ECR source at GSI with a permanent hexapole of 1~T at the pole tip.
The longitudinal magnetic field is 1~T at maximum as well. It has an electron
suppressing electrode at a potential of -2~kV. The extraction voltage is set to obtain
2.5~keV/u after extraction~\cite{Tinschert_private}.}
\label{fig_ecr_source}
\end{figure}

\begin{figure}
\centering
\epsfig{file=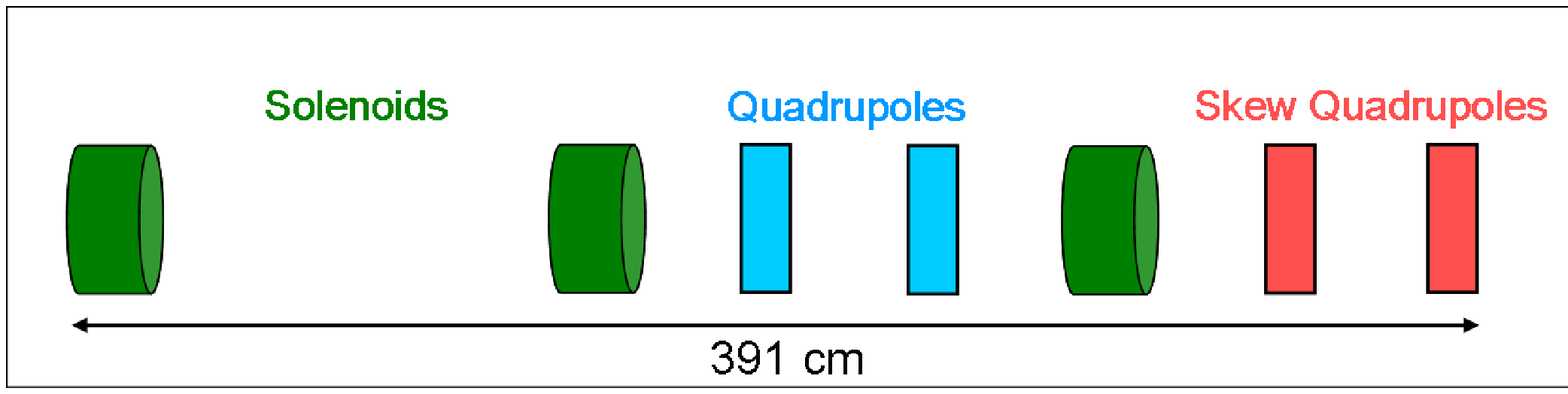,width=8.6cm,clip=}
\caption{(Color) Beam line from left to right for emittance transfer of a beam
extracted from an ECR source (here the GSI HLI source). It comprises three solenoids,
two quadrupoles, and two skew quadrupoles. Apertures are not to scale.}
\label{fig_ecr_line}
\end{figure}

\begin{figure}
\centering
\epsfig{file=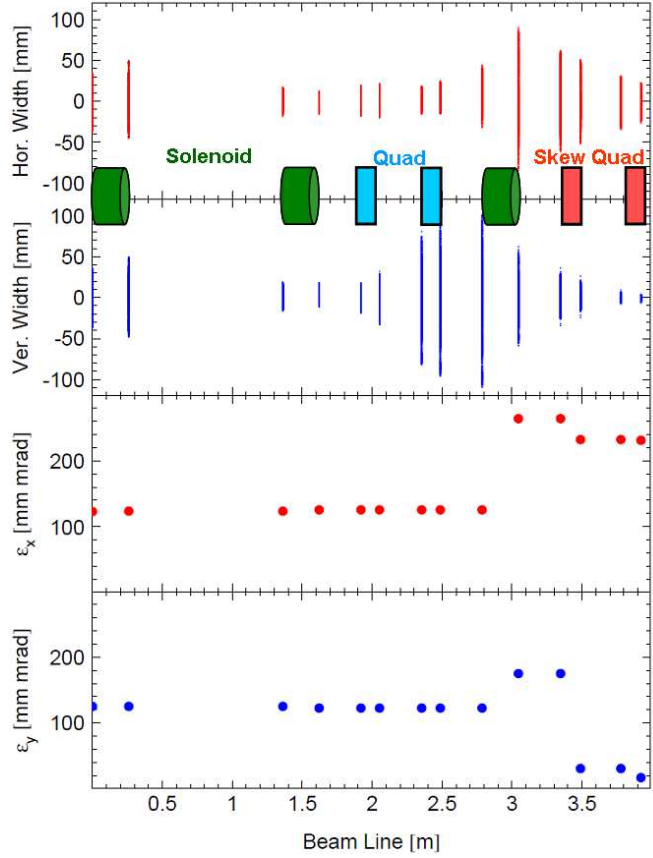,width=8.6cm,clip=}
\caption{(Color) From top to bottom: horizontal beam envelope,
vertical beam envelope, horizontal rms emittance, and vertical rms emittance
along the emittance transfer section for the ECR distribution.}
\label{fig_envelope_ecr}
\end{figure}

\begin{figure}
\centering
\epsfig{file=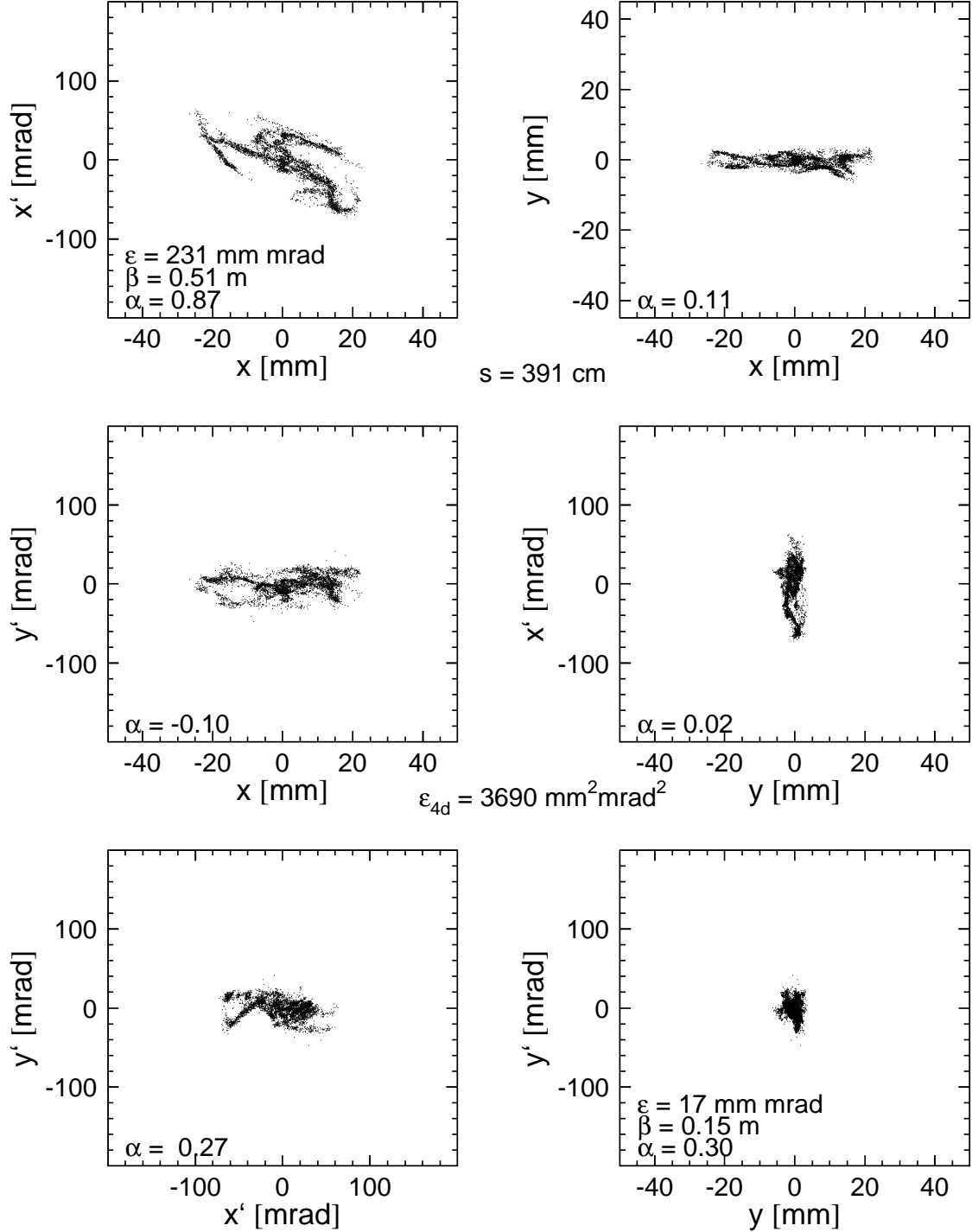,width=15cm,clip=}
\caption{Two dimensional projections of the ECR transverse phase space
distribution at the exit of the emittance transfer section. Horizontal
and vertical rms Twiss parameters are indicated as well as the four
inter-plane correlation parameters.}
\label{fig_tphell_ecr_s14}
\end{figure}

Round to flat transformation of ion beams extracted from an ECR source applies the same
concept as this transformation of electron
beams~\cite{PhysRevSTAB.4.053501,MOB13_LINAC2000}. For ions it was proposed
in~\cite{Bertrand_EPAC2006} assuming a beam second moment matrix $C$ with
non-zero elements just along its two diagonals. But generally all coupling moments
are different from zero and their magnitudes are similar. 
For electron and for ion beam extraction the required
non-symplectic transformation $M_{ns}$ is provided through placing the beam generation
inside a longitudinal field region. For electrons a solenoid is installed around the
cathode. In case of ions the longitudinal field for plasma confinement is an intrinsic
part of the ECR source. Prior to extraction the beam inside the plasma chamber is round
($\epsilon_x = \epsilon_y$) and is not correlated.
After extraction along the solenoidal fringe field the transverse phase space
distribution from the source has inter-plane correlations. Fig.~\ref{fig_tphell_ecr_s1}
shows such a distribution for \Oo ~extracted from the
ECR source at GSI (Fig.~\ref{fig_ecr_source}) with an energy of
2.5~keV/u~\cite{Spaedtke_private,SPAEDTKE_IPAC2010}.
\\
The distribution is obtained from multi-particle simulations and beam measurements
performed at the HLI are in good agreement with these
simulations~\cite{SPAEDTKE_IPAC2010,Spaedtke_LINAC2008}.
In Fig.~\ref{fig_tphell_ecr_s1} (and in the subsequent ones of same type)
correlations between two coordinates $u$ and $v$ are indicated by the dimensionless
$\alpha_{uv}$-parameter defined through the moments as
\begin{equation}
\label{def_alpha}
\alpha_{uv}\,:=\,\frac{-<uv>}{\sqrt{<u^2><v^2>-<uv>^2}}\,\,.
\end{equation}
The 120\degs ~symmetry in the inter-plane projections are due to the non-linear
inter-plane coupling hexapolar field
inside the ECR source. The observed inter-plane correlations
$\alpha_{uv}$ are from the solenoidal fringe field and the signs of the beam
moments \mbox{$<\!x'y\!>$} and \mbox{$<\!xy'\!>$} differ. Although the absolute values
of the inter-plane
$\alpha$-parameters seem small, the distribution is considerable correlated as seen
by comparing the product of the two transverse emittances with the four
dimensional rms emittance (see Eqs.~\ref{eps_4d} and~\ref{eps4d_biggerorequal}), i.e.
$\epsilon_x\cdot \epsilon_y\,=\,4.2\,\epsilon_{4d}$.
Due to the intrinsic inter-plane correlations the product
$\epsilon_x\cdot\epsilon_y$ exceeds the value required from four dimensional emittance
preservation by a factor of about four. That implies in turn that decreasing
$\epsilon_x\cdot\epsilon_y$ by this factor can be achieved by just removing
the inter-plane correlations from the beam using linear beam line elements.
\\
Figure~\ref{fig_ecr_line} displays a beam line that removes the inter-plane
correlations
from the beam. It comprises three solenoids, two quadrupoles, and two skew quadrupoles.
Its total length is about 4~m, the focusing strength are less than
0.5~m$^{\mathrm{-1}}$ (normal and skew quadrupoles) and 4.5~m$^{\mathrm{-1}}$
(solenoids).
After extraction from the source the beam is divergent in both transverse planes. To
provide
simultaneous focusing in both planes a solenoid is used as a first element of the beam
line. The beam line needs to remove the four inter-plane correlation moments, minimize
the vertical emittance (for instance) at its exit, and provide for beam envelopes
within a reasonable
aperture. In order to meet these boundaries simultaneously, several beam line elements
are needed and we found that the beam line in Fig.~\ref{fig_ecr_line} can do so.
Finding
the required focusing strengths is accomplished by numerical tools based on linear
transport of the second beam moments~(Eq.~\ref{def_transp_moments}). As input the
second moments extracted from the
distribution shown in Fig.~\ref{fig_tphell_ecr_s1} has been used. Knowing the focusing
strengths, multi-particle simulations using PARMTRA~\cite{PARMTRA} have been done.
Finite magnets
with hard edges were used. The energy spread of the beam is very low but has been
considered anyway. Space charge is of no concern at the HLI ECR source.
\\
Simulated beam envelopes together with transverse rms emittances along the beam line
are plotted in Fig.~\ref{fig_envelope_ecr}.
The beam is round (equal emittances) until the exit of the third solenoid. At this
location the distribution has maximum correlation with
$\epsilon_x\cdot\epsilon_y\,=\,12.5\,\epsilon_{4d}$. Despite this high value the
beam is not subject to losses. The emittances are dominated by wide angular
spreads. Within the final skew quadrupole the correlations are almost
completely removed. The distribution at the exit of the section is shown in
Fig.~\ref{fig_tphell_ecr_s14}.
Its vertical rms emittance is about 13 times
smaller than the horizontal one. Compared to its initial value the vertical rms
emittance was decreased by 86\% while the horizontal one increased by 85\%.
The product of the two transverse emittances was reduced from $4.2\,\epsilon_{4d}$
to $1.1\,\epsilon_{4d}$ while $\epsilon_{4d}$ was kept constant. The
remaining factor of 1.1 as well as residual values of the inter-plane
$\alpha$-parameters shows that the correlations were not completely removed. We
attribute this to non-linearities of the hexapolar coupling elements which cannot be
fully removed by linear elements. Reducing correlations using hexapoles is
proposed in~\cite{SPAEDTKE_IPAC2010}.

\section{Beams to be stripped}
\label{str_beam}
\begin{figure}
\centering
\epsfig{file=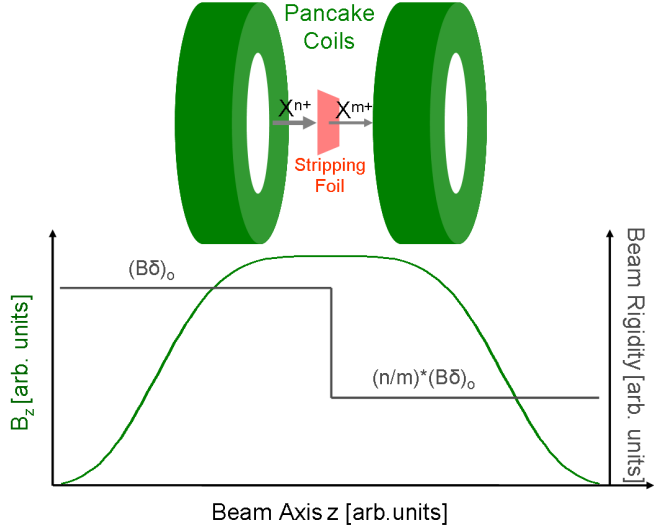,width=8.6cm,clip=}
\caption{(Color) Conceptual layout of an ion stripper set-up
providing a non-symplectic transformation $M_{ns}$ for later transverse emittance
transfer. The stripping foil is placed between two pancake coils, i.e. inside a
region with a longitudinal magnetic field of few Tesla in strength.}
\label{fig_solstrip}
\end{figure}

\begin{figure}
\centering
\epsfig{file=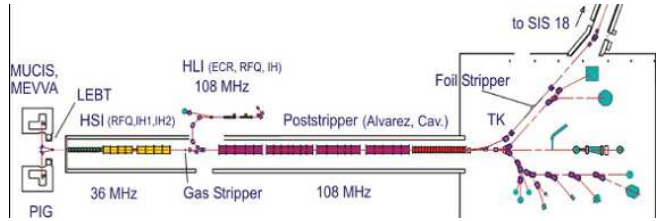,width=8.6cm,clip=}
\caption{(Color) The Universal Linear Accelerator (UNILAC) at GSI.}
\label{fig_UNILAC}
\end{figure}

\begin{figure}
\centering
\epsfig{file=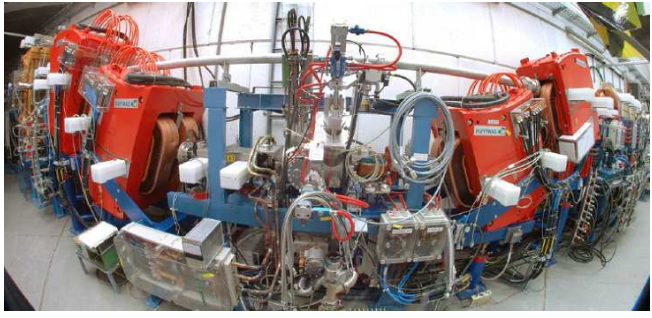,width=8.6cm,clip=}
\caption{(Color) Fish eye photograph of the foil charge state stripper at the GSI
UNILAC including the four bend analyzing chicane. The beam enters from the left.}
\label{fig_LASEP}
\end{figure}

\begin{figure}
\centering
\epsfig{file=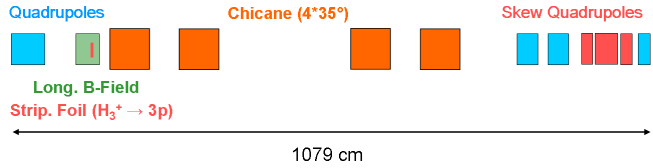,width=8.6cm,clip=}
\caption{(Color) Beam line from left to right for emittance transfer of a beam along
the UNILAC. It comprises two quadrupole singulets, a stripping foil in between two
pancake coils, a four bend chicane, a doublet, and a skew quadrupole triplet.
Apertures are not to scale.}
\label{fig_str_line}
\end{figure}

\begin{figure}
\centering
\epsfig{file=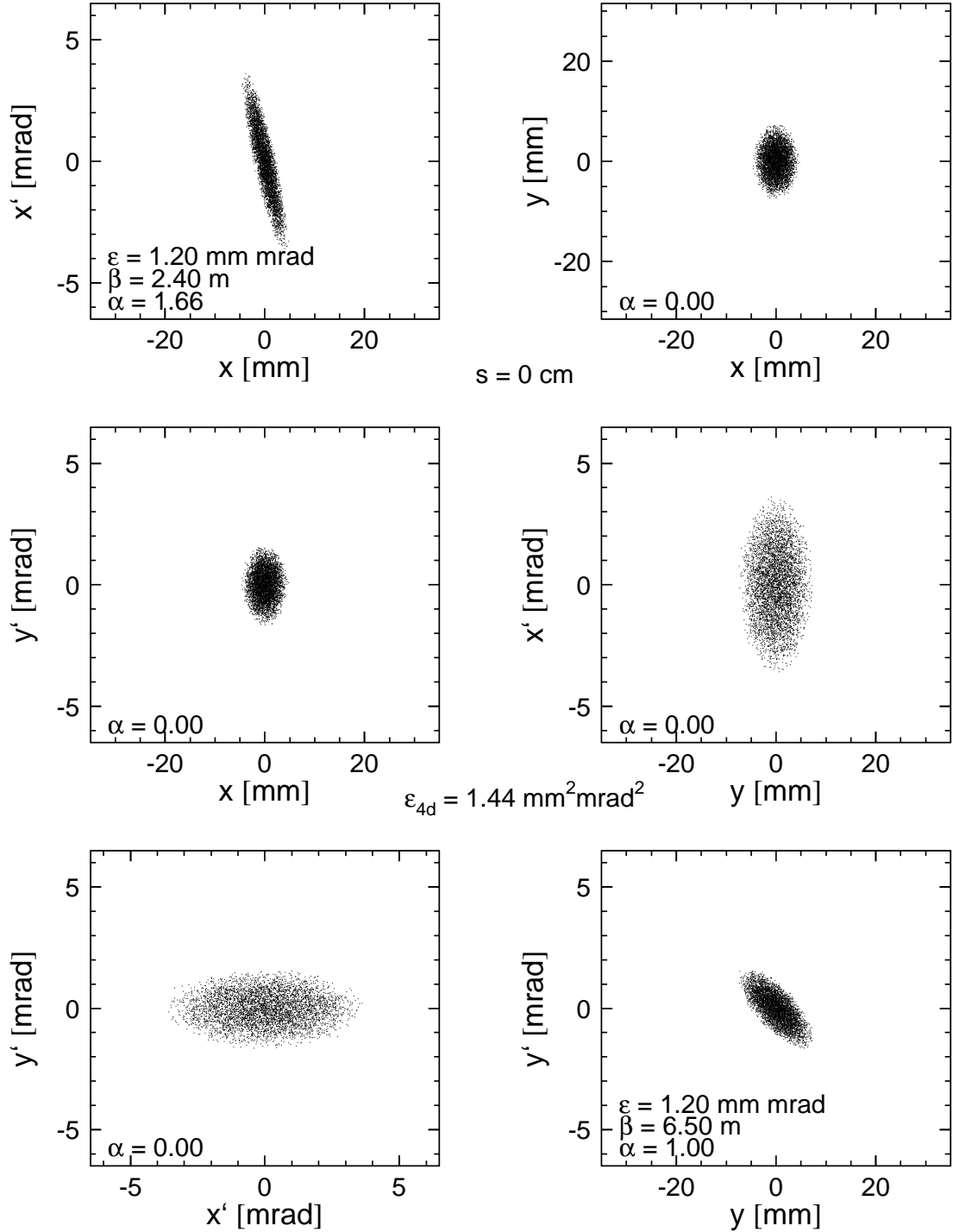,width=15cm,clip=}
\caption{Two dimensional projections of the transverse phase space
distribution at the entrance to the proposed emittance transfer section along the
UNILAC. Horizontal and vertical rms Twiss parameters are indicated as well as the four
inter-plane correlation parameters.}
\label{fig_tphell_str_start}
\end{figure}

\begin{figure}
\centering
\epsfig{file=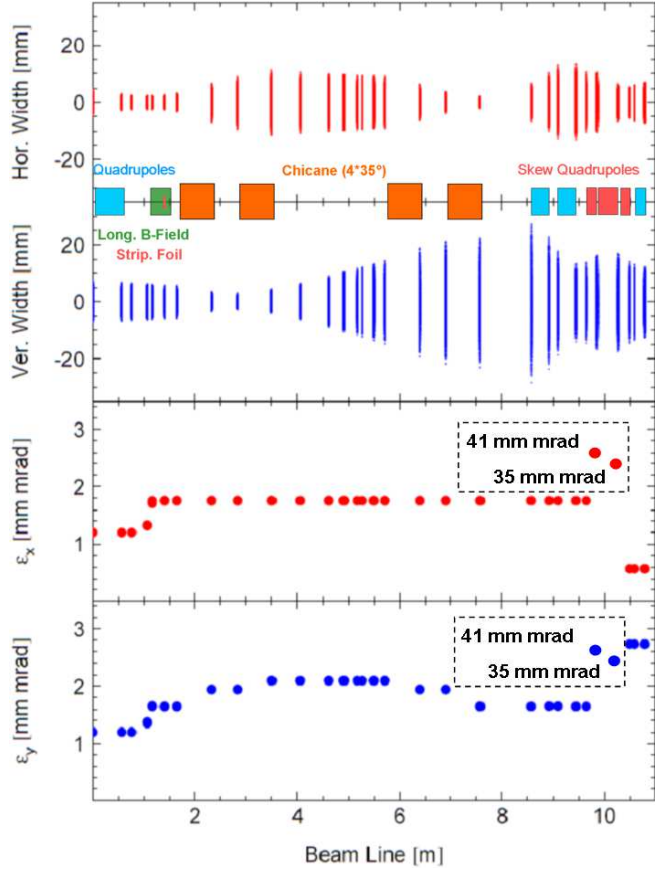,width=8.6cm,clip=}
\caption{(Color) From top to bottom: horizontal beam envelope,
vertical beam envelope, horizontal rms emittance, and vertical rms emittance
along the proposed emittance transfer section along the UNILAC.}
\label{fig_envelope_str}
\end{figure}

\begin{figure}
\centering
\epsfig{file=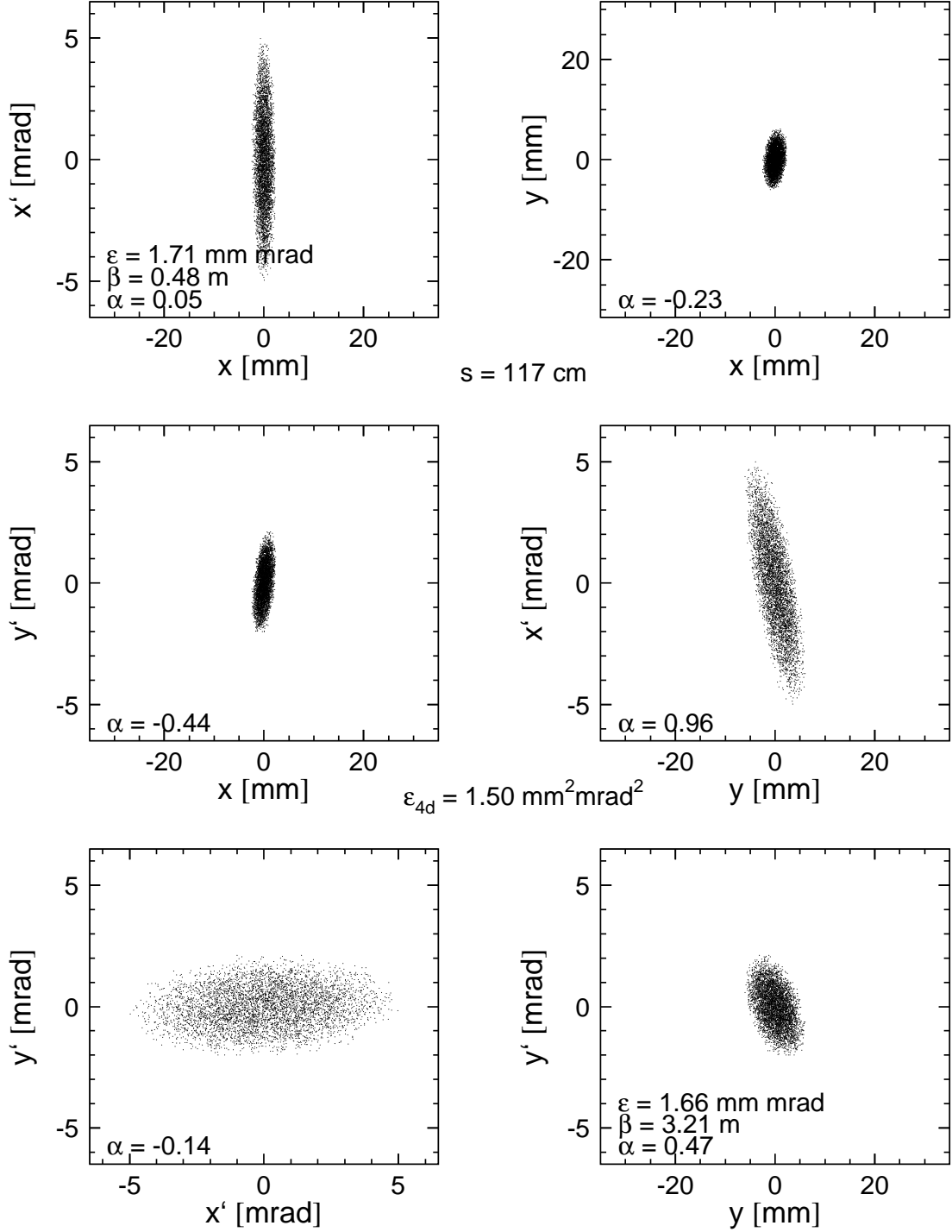,width=15cm,clip=}
\caption{Two dimensional projections of the transverse phase space
distribution at the exit of the longitudinal magnetic field section of the proposed
emittance transfer section along the UNILAC.
Horizontal and vertical rms Twiss parameters are indicated as well as the
four inter-plane correlation parameters.}
\label{fig_tphell_str_solexit}
\end{figure}

\begin{figure}
\centering
\epsfig{file=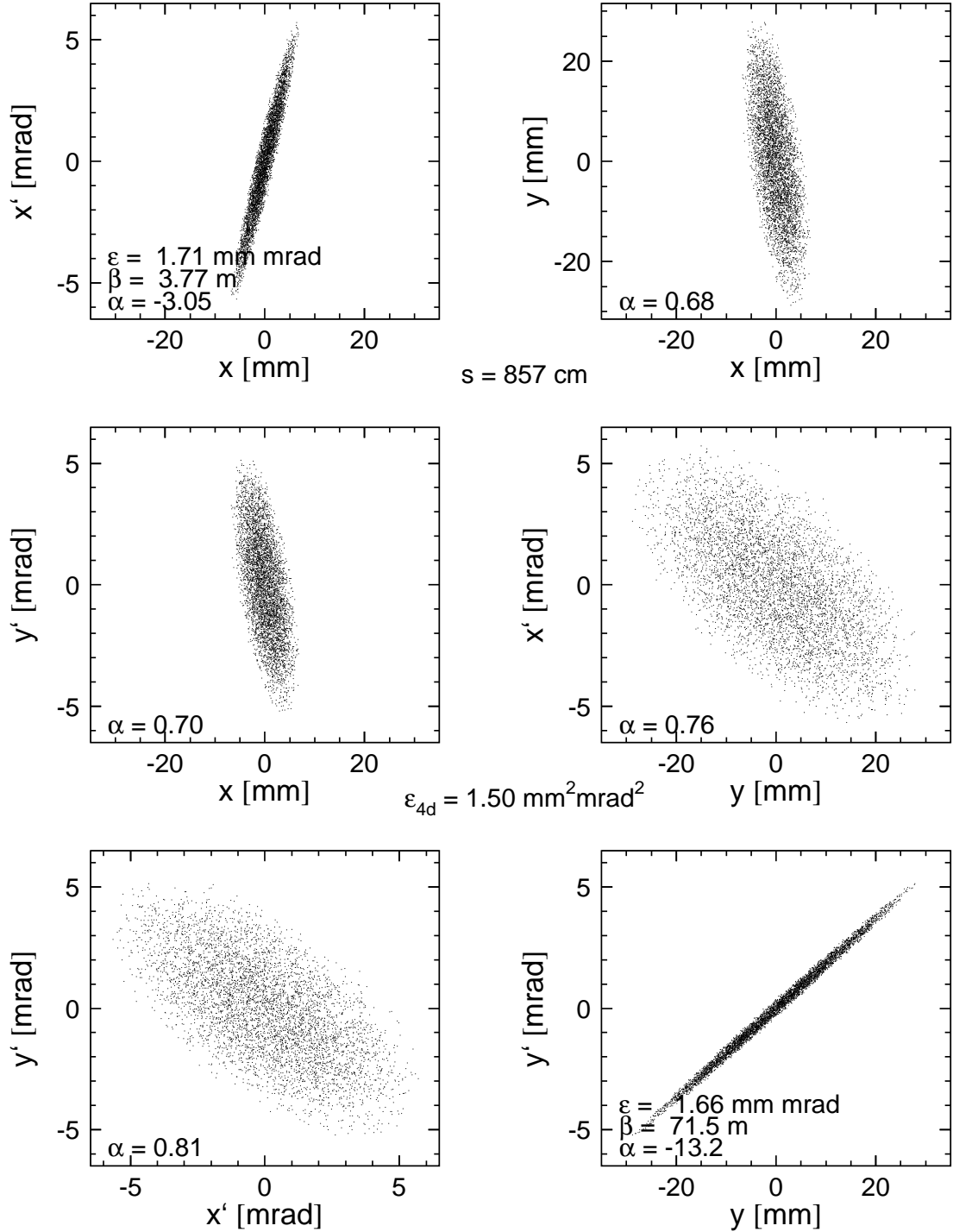,width=15cm,clip=}
\caption{Two dimensional projections of the transverse phase space
distribution at the entrance to the first quadrupole after the chicane of the proposed
emittance transfer section along the UNILAC.
Horizontal and vertical rms Twiss parameters are indicated as well as the
four inter-plane correlation parameters.}
\label{fig_tphell_str_tk3qd6in}
\end{figure}

\begin{figure}
\centering
\epsfig{file=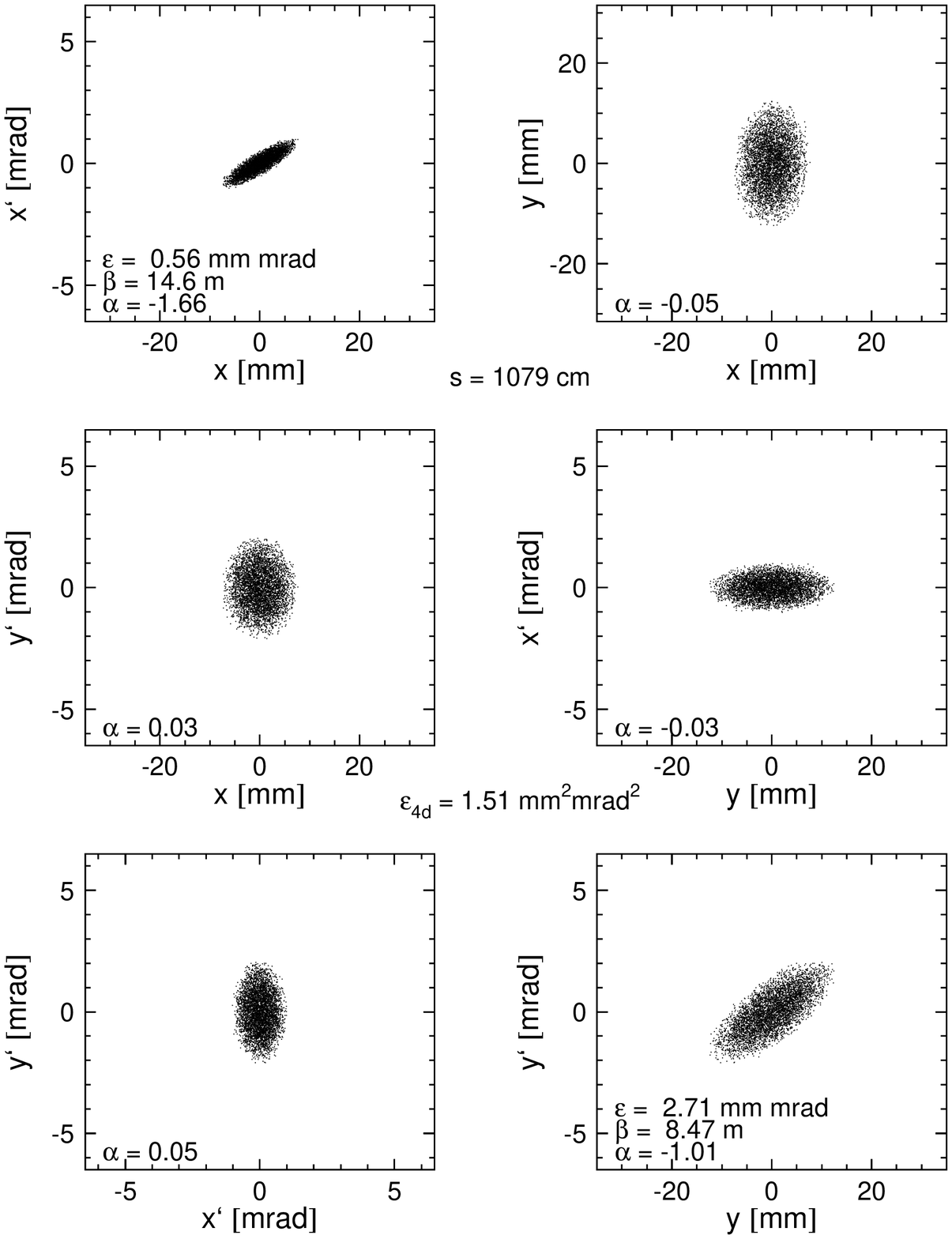,width=15cm,clip=}
\caption{Two dimensional projections of the transverse phase space
distribution at the exit of the proposed emittance transfer section along the UNILAC.
Horizontal
and vertical rms Twiss parameters are indicated as well as the four
inter-plane correlation parameters.}
\label{fig_tphell_str_s56}
\end{figure}

ECR sources generally do not provide high beam intensities for heavy and intermediate
mass ions. Other source types are used in these cases (Multi-Cusp, Metal Vapor Vacuum
Arc), which provide low
charge states but high particle currents. From first principles also
such sources can be designed applying ion extraction along a solenoidal
fringe field to allow for later round to flat transformation. However, at extraction
energy high intensity beam dynamics is subject to space charge forces.
Latest at the entrance to an Radio Frequency Quadrupole (RFQ) eventual space charge
neutralization does not mitigate any longer the effect of beam self fields.
Acceleration and transport with space charge might strongly reduce or
remove the required inter-plane correlation that were imposed previously at
extraction along a solenoidal fringe field. For this reason the beam should be exposed
to the fringe field after some acceleration, when space charge forces are
sufficiently reduced. On the other hand ion beams from low charge state, high
particle current sources are stripped after gaining some energy in order to keep the
acceleration process reasonable efficient.
\\
The obstacle to meet is to provide for a non-symplectic
transformation $M_{ns}$ after acceleration. Unlike in the ion source, the accelerated
beam must enter and leave the solenoidal field and a complete solenoid is a symplectic
transformation which cannot be used to prepare round to flat transformation.
\\
The transformation through the solenoid is non-symplectic if the beam rigidity is
abruptly changed in between the entrance and exit fringe fields, i.e. if the beam
properties are reset inside the solenoid. In this case the
focusing strengths $K$ from Eq.~\ref{SolFringe} are different at entrance and exit and
the beam transport through the solenoid is a linear non-symplectic transformation
$M_{ns}$ preserving~$\epsilon_{4d}$.
\\
We propose to merge the required charge stripping of ions from low charge state,
high particle current sources and the provision of a transformation $M_{ns}$ for
transverse emittance transfer preparation by placing the charge stripping foil in
between two pancake coils as drawn in~Fig.~\ref{fig_solstrip}.
Stripping abruptly
changes the beam rigidity and has to be performed anyway along the ion linac. Based on
the
existing charge stripping section of the GSI UNILAC a beam line was conceptually
designed that can provide horizontal to vertical emittance transfer.
\\
The UNILAC (Fig.~\ref{fig_UNILAC}) comprises three source terminals, two
Drift Tube Linac (DTL) sections
separated by a gaseous stripper section, and a transfer channel to the synchrotron
SIS18. Along the DTL sections the beam is accelerated to 1.4~MeV/u and 11.4~MeV/u,
respectively. The transport channel to the synchrotron
includes a foil stripper section shown in the photograph of Fig.~\ref{fig_LASEP}.
The foil stripper section~\cite{LASEP} comprises
a quadrupole singulet, a stripping foil, a vertical four bend chicane, and a
quadrupole doublet. The chicane provides dispersion in its center for charge state
separation. To demonstrate emittance transfer the beam line can be modified as
drawn schematically in Fig.~\ref{fig_str_line}.
GSI aims at experimental verification of the emittance transfer concept. The
modifications of the existing foil stripper beam line will
be conceived such that they do not intercept the beam in case no emittance
transfer is required. The effective length of the longitudinal magnetic field region is
40~cm and the foil is placed 10~cm from the exit of this region.
First beam tests are to be performed using a beam of
\Htp ~stripped to a proton beam in a 20~$\mu g/cm^{\mathrm{2}}$ carbon foil placed
between to pancake coils. Protons were chosen in order to simplify the first proof of
principles: low focusing strengths allow for usage of on-site equipment, low beam
current avoids space charge effects, and a single charge state spectrum after
stripping is provided. Multi-particle simulations for this case have been done and are
presented in the following.
\\
The change of charge state, i.e. rigidity, inside a longitudinal magnetic field region
is modeled by matrix
transformation of single particle coordinates according to
Eq.~\ref{transport_matrix}. The matrix includes to parts $M_{Sol_i}$ and $M_{Sol_o}$;
the first part comprising a solenoid entrance fringe field and the longitudinal
magnetic field until the stripping foil:
\begin{equation}
\label{M_Sol_i}
M_{Sol_i}\,=\,
\begin{bmatrix}1+\frac{KL(1-C)}{\alpha} & \frac{SL}{\alpha} & \frac{SLK}{\alpha} & -\frac{L(1-C)}{\alpha}\\
KS & C & CK & -S\\
-\frac{SLK}{\alpha} & \frac{L(1-C)}{\alpha} & 1+\frac{KL(1-C)}{\alpha} & \frac{SL}{\alpha}\\
-CK & S & SK & C\\
\end{bmatrix}
\end{equation}
with
\begin{equation}
C\,:=\,cos(KL),\,\,\,\,\,S\,:=\,sin(KL),\,\,\,\,\,\alpha\,:=\,2\,KL\,,
\end{equation}
$K$ from Eq.~\ref{SolFringe}, and $L$ as distance from the effective longitudinal
field edge to the foil.
The beam rigidity is calculated from the unstripped charge state, i.e. using \Htp ~at
11.4~MeV/u. The foil itself is modeled by increasing the spread of the angular
distribution through scattering, i.e.
\begin{equation}
<x'^2>+ <y'^2>\,\longrightarrow\,<x'^2>+ <y'^2>\,+\,\Delta \varphi^2\,.
\end{equation}
The amount of scattering $\Delta \varphi$ was calculated using
the ATIMA code~\cite{ATIMA} as 0.296~mrad. Due to scattering the four dimensional
emittance $\epsilon_{4d}$ increases during the stripping process. Transport from
the foil to the exit of the longitudinal magnetic field region is modeled by a matrix
corresponding to $M_{Sol_i}$, i.e.
\begin{equation}
\label{M_Sol_o}
M_{Sol_o}\,=\,
\begin{bmatrix}1 & \frac{SL}{\alpha} & 0 & -\frac{L(1-C)}{\alpha} \\
0 & C-\frac{KL(1-C)}{\alpha} & -K & -S-\frac{SLK}{\alpha}\\
0 & \frac{L(1-C)}{\alpha} & 1 & \frac{SL}{\alpha}\\
K & \frac{SLK}{\alpha}+S & 0 & C-\frac{KL(1-C)}{\alpha}\\
\end{bmatrix}
\end{equation}
using the beam rigidity corresponding to the stripped beam, i.e. protons at 11.4~MeV/u.
Here $L$ is the distance from the foil to the exit of the longitudinal magnetic field
region.
According to~\cite{ATIMA} the energy loss and straggling in the foil can be
neglected. It must be emphasized that the increase of $\epsilon_{4d}$ is
purely from scattering in the foil; it is not caused by the change of beam rigidity
inside the longitudinal field. Scattering in the foil occurs anyway along beam
lines that include foil strippers. The simulations additionally included beam momentum
spread, hexapolar fringe fields of the chicane dipoles, and the existing aperture
limitations.
\\
Simulations started at the entrance to the quadrupole singulet in front of
the chicane. Fig.~\ref{fig_tphell_str_start} shows the initial
transverse phase space distribution.
Initial six dimensional Twiss parameters were concluded from various beam experiments
with \Htp ~beams. Beam spot images from fluorescence screens close this position did
not reveal any x-y correlation. Therefore initial inter-plane correlations are assumed
as zero. The distribution was tracked through the emittance transfer section and
corresponding envelopes and transverse emittances are plotted in
Fig.~\ref{fig_envelope_str}.
Full beam transmission is achieved.
The observed increase in both transverse emittances at the entrance to the
longitudinal magnetic field region is driven by inter-plane correlations caused by
the fringe field. There is no increase of the four dimensional emittance
$\epsilon_{4d}$ until the stripping foil. However, the increase of $\epsilon_{4d}$
at the foil from scattering is just 4\%. Vertical dispersion from the first dipole
causes additional increase of the vertical and the four dimensional emittance, both
being compensated later by the last dipole.
\\
The distribution at the exit of the longitudinal magnetic field region is shown in
Fig.~\ref{fig_tphell_str_solexit}.
It has inter-plane correlations and the signs
of \mbox{$<\!x'y\!>$} and \mbox{$<\!xy'\!>$} differ as expected from the action of
the solenoidal fringe fields. The product of the transverse emittances is a factor
of 1.8 larger than the four dimensional emittance~$\epsilon_{4d}$. During the transport
through the chicane the horizontal emittance is preserved. The intrinsic inter-plane
correlation of the beam, i.e. the ratio $\epsilon_x\epsilon_y/\epsilon_{4d}$
(corrected w.r.t. vertical dispersion) remains constant. Along the elements
after the chicane the
correlations are removed, the horizontal emittance is minimized, and the beam is
re-matched for further transport to the synchrotron. To accomplish these tasks within
the existing apertures, the gradients of the quadrupole doublet, the skew triplet, and
the final singulet are determined numerically. The same algorithm as mentioned
in Sec.~\ref{ecr_beam} is applied. As input for the algorithm the second beam moments
extracted from the distribution at the doublet entrance are
used~(Fig.~\ref{fig_tphell_str_tk3qd6in}).
Here we presently rely on the simulations using the PARMTRA
code being part of the PARMILA code family~\cite{PARMILA}. The later has been
benchmarked successfully with beam experiments involving more complex beam lines even
with strong space charge~\cite{PhysRevSTAB.11.094201,PhysRevLett.102.234801}.
The beam moments at the doublet entrance also can be measured if a pepper-pot device
will be installed right behind the last singulet. Its installation is planned for the
experimental verification of the proposed emittance transfer concept.
\\
After applying the required gradients for emittance transfer and horizontal emittance
minimization the beam is transported until the exit of the last singulet. The resulting
distribution is plotted in Fig.~\ref{fig_tphell_str_s56}.
Inter-plane correlations are almost completely removed. Residual values are due to
hexapolar dipole fringe fields as well as to chromaticity in quadrupoles and
skew quadrupoles from the finite beam momentum spread~($\approx$~0.1\%~rms).
The product of
$\epsilon_x\cdot\epsilon_y$ exceeds $\epsilon_{4d}$ by less than one percent.
Starting from an initial transverse emittance ratio of 1.0, the final ratio is~4.8.
The horizontal emittance is reduced by~53\% and the vertical one increased
by~125\%. The four dimensional emittance $\epsilon_{4d}$ increased by~5\% mainly
due to angular scattering in the stripper foil~(4\%). The remaining part is from
beam line non-linearities as mentioned before.
\\
Unlike for the round to flat transformation of beams from an ECR source
(Sec.~\ref{ecr_beam}), the longitudinal field strength at the stripper foil is a
free parameter. The amount of emittance transfer scales with the field strength.
However, practical limitations for this amount rise from finite apertures.

\section{Conclusion and Outlook}
It was shown that flat to round transformation is not restricted to electron beams
but can be accomplished for beams of ions as well. The required
solenoid fringe fields are provided intrinsically for beams being extracted from an
ECR source. Beams from other sources can be emittance-shaped by applying charge state
stripping inside a longitudinal magnetic field. Examples for the layout of emittance
transfer beam lines were presented for both cases. Multi-particle simulations
demonstrated that the transfer is feasible. An initial round beam could be transformed
into a beam with transverse emittance ratio of almost five. The amount of transfer for
stripped ions can be controlled through the longitudinal magnetic field strength.
\\
GSI aims at experimental verification of emittance transfer by assembling the beam
lines presented in this paper. Emittance transfer of beams from ECR sources might by
of special interest for facilities operating an ECR source in connection with
multi-turn injection into a synchrotron, as cancer treatment facilities for
instance~\cite{HIT}. High intensity beams of intermediate or heavy mass ions
for synchrotrons are to be provided for the FAIR project~\cite{FAIR}. Stripping of
those ions results in a charge state spectrum and requires charge state separation.
Additionally, stripping of heavy ions uses much thicker foils
($\approx$~600~$\mu g/cm^{\mathrm{2}}$) with respect to light ions, thus the four
dimensional emittance growth from scattering is significantly higher~($\geq$~70\%).
This growth reduces the budget for the amount of emittance transfer. However, emittance
transfer lines can be designed also for such scenarios although the layout
will be more complex than the one for protons presented here.
Tentative layouts were made on the
emittance transfer for beams of \Ul ~stripped to \Uh ~based on the existing foil
stripping section at the GSI UNILAC. Horizontal emittance reduction of up to~40\% was
achieved including space charge and charge state separation. Further optimization of
the layout for an emittance transfer section for intense beams of heavy ions is
needed. This optimization is beyond the scope of the present paper.

\begin{acknowledgments}
The author wishes to express his gratitude to L.~Dahl, S.~Ratschow, and the 
ECR source team at GSI for fruitful discussions. 
\end{acknowledgments}

\bibliography{groening}

\end{document}